\documentclass[12pt]{amsart}

\usepackage{amssymb}
\usepackage{upref}

\theoremstyle{definition}

\theoremstyle{remark}

\newcommand{\Proof}{\begin{proof}} 

\def\ldots{\mathinner{\ldotp\ldotp\ldotp}}
\def\ldots{\mathinner{\cdotp\cdotp\cdotp}}

\def\R{\mathbb R}

\def\D{\mathcal D}

\numberwithin{equation}{section}
\allowdisplaybreaks
\begin{document}

\title[The form boundedness 
criterion]{\bf The form boundedness 
criterion  \\ for 
the relativistic Schr\"odinger operator}

\author[V.~G. Maz'ya]
{V.~G. Maz'ya}
\address{Department of Mathematics,
Link\"oping University, SE-581 83, Link\"oping, 
Sweden}
\email{vlmaz@mai.liu.se}
\thanks{$^*$Supported in part 
by NSF Grant DMS-0070623.}
\author[I.~E. Verbitsky]
{I.~E. Verbitsky$^*$}
\address{Department of Mathematics,
University of Missouri,
Columbia, MO 65211, USA}
\email{igor@math.missouri.edu}

\begin{abstract} 
 We establish necessary and sufficient conditions 
for the boundedness of the relativistic Schr\"odinger 
operator 
$\mathcal{H} = \sqrt{-\Delta} + Q$ 
from the Sobolev space $W^{1/2}_2 (\R^n)$ to its dual 
$W^{-1/2}_2 (\R^n)$,  for 
an arbitrary real- or complex-valued potential $Q$ on $\R^n$. 
In other words, we give a complete solution to the problem 
of the domination of the potential energy  
 by the kinetic
energy  in the relativistic case characterized by the 
inequality
$$ 
\left \vert \int_{\R^n} |u(x)|^2 \, Q(x) \, dx\right 
\vert \leq \text{const} \, ||u||^2_{W_2^{1/2}}, \quad u 
\in C^\infty_0(\R^n),
$$
where the ``indefinite weight'' $Q$ is a locally integrable function 
(or, more generally,  a
distribution)  on $\R^n$.  Along with necessary and sufficient 
results, we also present new  broad classes of admissible potentials $Q$ 
in the scale of Morrey spaces of negative order,
and discuss their relationship  to well-known $L_p$ and Fefferman-Phong
conditions. \end{abstract}


\maketitle

\vskip20pt 

\section{Introduction}

In the present paper we establish necessary and sufficient conditions 
for the 
 relative form
boundedness  of the potential energy operator $Q$ with respect to the 
{\it relativistic\/} kinetic energy operator 
$\mathcal{H}_0 = \sqrt{-\Delta}$, which is fundamental 
to relativistic quantum systems. Here $Q$ is an  arbitrary real- or
complex-valued  potential (possibly a distribution), and $\mathcal{H}_0$ 
is a nonlocal  operator which replaces the standard Laplacian 
$H_0=-\Delta$ used  in the  nonrelativistic theory. 

More precisely, we characterize all potentials $Q \in \D'(\R^n)$ 
such that 
$$
|\langle Q \, u, \,u \rangle |
 \le a \,  \langle \sqrt{-\Delta} \, u, \, u\rangle  + 
b \,  \langle  u, \,u \rangle,
\qquad u \in \D(\R^n), 
$$
for some $a>0, \, b\in\R$. 

In particular, if $Q$ is real-valued, 
and  the form bound $a<1$, then  
 this inequality  makes it possible 
to define, via the classical KLMN Theorem (see, e.g., \cite{RS}, 
Theorem X.17),
the relativistic Schr\"odinger operator 
$\mathcal{H}=\sqrt{-\Delta} + Q$, where the sum $\sqrt{-\Delta} + Q$ is 
a uniquely defined self-adjoint operator 
associated with the sum
of  the corresponding quadratic forms whose form domain
$\mathcal{Q}(\mathcal{H})$ 
coincides with the Sobolev space $W_2^{1/2}(\R^n)$.  
(For complex-valued $Q$, 
this sum   defines an  $m$-sectorial operator provided  $a< 1/2$; 
see \cite{EE}, Theorem IV.4.2.) 

Equivalently, we give a complete characterization of the class of 
admissible potentials $Q$ such that 
 the  relativistic Schr\"odinger 
operator 
$\mathcal{H} = \sqrt{-\Delta} + Q$ is bounded 
from  $W^{1/2}_2 (\R^n)$ to the dual space 
$W^{-1/2}_2 (\R^n)$.

A nice introduction to the theory 
 of the relativistic Schr\"odinger operator is given in \cite{LL}. 
We  observe that it is customary to develop the relativistic 
theory in parallel to its nonrelativistic counterpart, without making a 
 connection between them. One of the advantages of our general approach   
where distributional potentials $Q$ are admissible is that it provides 
a direct link between the two theories. 

In Sec. 2, we develop an extension principle   which establishes a 
connection between the 
relativistic  Schr\"odinger operator $\mathcal{H} = \sqrt{-\Delta} + Q$
 and the  nonrelativistic one, $H = - \Delta +
\widetilde{Q}$, where $\widetilde{Q}$ is a distribution   defined on
a higher dimensional  Euclidean space. Note that the 
{\it nonrelativistic\/}  form boundedness problem 
was settled in full generality  only recently 
by the authors in \cite{MV2}. (The one-dimensional case of the 
Sturm-Liouville  operator $H=- \frac {d^2} {dx^2} + Q$ on the real axis 
and half-axis is treated 
in \cite{MV3}.)

It is worth noting  that in the above discussion of the relative form
boundedness   $\mathcal{H}_0 = \sqrt{-\Delta}$ can 
be replaced by 
 $\mathcal{H}_{\mathfrak m} =  
 \sqrt{-\Delta + {\mathfrak m}^2} - {\mathfrak m}$, 
where 
$\mathfrak m$ represents the mass of the particle under consideration. 
This operator appears in the relativistic Schr\"odinger 
equation:
\begin{equation}\label{E:1.1}
\mathcal{H}_{\mathfrak m}  \psi + Q \, \psi = E  \,
 \psi \quad
\text{in}  \quad \D',
\end{equation}
where $\D = C^\infty_0(\R^n)$.

\vskip13pt

One of the central questions of the relativistic theory is the 
domination of the  potential energy 
$\int_{\R^n} |u|^2 \, Q(x) \, dx$
 by the kinetic energy associated with 
$||u||^2_{W_2^{1/2}}$, which 
explains a special role of  the Sobolev space 
$W_2^{1/2}$ in this context (see \cite{LL}, Sec. 7.11 and 11.3).  
We address this problem by characterizing the weighted norm  
inequality  with
``indefinite weights'': \begin{equation}\label{E:1.2}
\left \vert \int_{\R^n} |u(x)|^2 \, Q(x) \, dx\right 
\vert \leq \text{const} \, ||u||^2_{W_2^{1/2}}, \qquad u 
\in \D. 
\end{equation}
Here 
$Q$ is
a locally integrable real- or
complex-valued function, or more generally,  a 
 distribution. In the 
latter case,  the  left-hand side of (\ref{E:1.2}) is understood 
as  
$\, |\langle Q \, u, \, u\rangle|$, where 
$\langle Q \cdot, \, \cdot\rangle $
is the quadratic form associated with the 
 corresponding multiplication
operator. 

An analogous inequality characterized in \cite{MV2}, 
\begin{equation}\label{E:1.3}
\left \vert \int_{\R^n} |u(x)|^2 \, Q(x) \, dx\right 
\vert \leq \text{const} \, ||u||^2_{W_2^{1}}, \qquad u 
\in \D,
\end{equation}
with the Sobolev norm of order $1$ in place of $1/2$, 
is used extensively in  spectral 
theory 
of the  nonrelativistic  
Schr\"odinger operator $H = - \Delta + Q$.
(See  \cite{AiS}, \cite{Fef}, \cite{M1}, \cite{M2}, \cite{MV2}, 
 \cite{Nel}, \cite{RS}, \cite{Sch},
\cite{Sim}.)   In particular,  (\ref{E:1.3}) is equivalent to 
the relative
form boundedness of  the potential energy operator 
$Q$ with respect to the traditional 
kinetic  energy operator $H_0 = - \Delta$.

We remark that, for {\it nonnegative\/} (or nonpositive) potentials $Q$
(possibly  measures on $\R^n$ which may be
singular with respect to $n$-dimensional Lebesgue measure),   
the inequalities
(\ref{E:1.2}) and (\ref{E:1.3}) have been thoroughly studied, and 
are well understood by now. (See \cite{ChWW}, \cite{Fef}, 
\cite{KeS}, \cite{M1}, \cite{MV1},  \cite{Ver}.) On the other hand, for
real-valued $Q$ which may change sign,   or complex-valued $Q$, only 
sufficient conditions, as well 
as examples of potentials with strong cancellation
properties have been known, mostly in the 
framework of the nonrelativistic Schr\"odinger operator theory  and 
Sobolev 
multipliers (\cite{AiS}, \cite{CoG}, \cite{MSh}, 
 \cite{Sim}). \vskip13pt

We now state our main results on the relativistic Schr\"odinger operator 
with ``indefinite'' potentials $Q$  in the
form of the following two  theorems. Simpler sufficient and necessary
conditions in the scales of Sobolev, Lorentz-Sobolev,  and  
Morrey spaces of
negative order are obtained as corollaries. Their relationship
to  more conventional $L_p$ and Fefferman-Phong classes is discussed 
at the end of the Introduction, and in Sec. 3 in more detail.  

 Note that rigorous definitions 
of the expressions like 
$\langle Q \cdot, \, \cdot \rangle$ 
or $(-\Delta +1)^{-1/4} Q$ 
 are given in the main body of the paper. 

\vskip13pt

\noindent {\bf Theorem I.} 
{\it 
Let $Q \in \mathcal \D'(\R^n)$, $n \ge 1$.
The following statements are equivalent: 

{\rm (i)} The relativistic Schr\"odinger operator $\mathcal{H} = 
\sqrt{- \Delta} + Q$
 is  bounded 
from  $W^{1/2}_2 (\R^n)$ to $W^{-1/2}_2 (\R^n)$.

{\rm (ii)} The inequality
\begin{equation}\label{E:1.4}
|\langle Q u, \, u\rangle | \ \leq \text{\rm{const}} 
\, ||u||^2_{W^{1/2}_2}, 
\qquad \forall u \in \D,
 \end{equation}
holds, where the constant does not depend on  $u$.

{\rm (iii)}  $\Phi = (-\Delta +1)^{-1/4} Q \in L_{2, \, loc}(\R^n)$, 
and the  inequality
\begin{equation}\label{E:1.5}
 \int_{\R^n} |u(x)|^2 \, |\Phi (x)|^2 \, dx \leq 
\text{\rm{const}} \, \, ||u||^2_{W^{1/2}_2}, \qquad \forall u \in \D,
\end{equation}
holds, where the constant does not depend on  $u$.}

\vskip13pt

\noindent {\bf Theorem II.} 
{\it Let  $Q \in \mathcal \D'(\R^n)$, $n \ge 1$, and let 
$\mathcal{H} = 
\sqrt{- \Delta} + Q$. Then $\mathcal{H}: \, 
W^{1/2}_2 (\R^n)\to W^{-1/2}_2 (\R^n)$ is bounded if and only 
if $\Phi = (-\Delta +1)^{-1/4} Q \in L_{2, \, loc}(\R^n)$, 
and any one of the following equivalent conditions holds:

{\rm (i)} For every compact set $e \subset \R^n$,  
\begin{equation}\label{E:1.6}
\int_e |\Phi (x)|^2 \, dx \le
\text{\rm{const}}  \,  \, \text{\rm{cap}} \, (e,  \, W^{1/2}_2),
\end{equation}
where the constant does not depend on $e$. Here $\text{\rm{cap}} \, 
(\cdot,  \,
W^{m}_2)$ is the capacity associated with the  Sobolev space
$W_2^m(\R^n)$  defined by: $$\text{\rm{cap}} \, (e, \,
W_2^m) = \inf \, \{ ||u||_{W_2^m}^2: \quad u \in \D, \qquad u
\ge 1 \quad \text{on} \quad e\}.$$

{\rm (ii)} The function 
  $J_{1/2} \, |\Phi|^2$ is finite 
$\text{\rm{a.e.}}$, and 

\begin{equation}\label{E:1.7} 
J_{1/2} \,  \left (J_{1/2} \, |\Phi|^2 \right)^2 (x) \le \text{\rm{const}}  
 \, J_{1/2} \, |\Phi|^2(x) \quad
\quad \text{\rm{a.e.}}
\end{equation}
Here $J_{1/2}= (-\Delta +1)^{-1/4}$ is the Bessel potential of 
order $1/2$. 

{\rm (iii)} For every dyadic
cube $P_0$ in  $\R^n$ of sidelength  $\ell(P_0) \le 1$, 
\begin{equation}\label{E:1.8}
\sum_{P \subseteq P_0} 
 \left [ \frac {\int_P |\Phi (x)|^2 \, dx }{
|P|^{1 - 1/(2n)}} \right ]^2 |P| \le  \text{\rm{const}}  \, 
 \int_{P_0} |\Phi (x)|^2 \, dx ,
\end{equation}
where the sum is taken over all dyadic cubes $P$ contained in $P_0$, 
and the constant does not depend on $P_0$.} 

\vskip13pt

We observe that statement (iii) of Theorem I reduces 
 the problem
of characterizing general weights $Q$ such that either (i) or equivalently 
(ii)  holds,  to a similar problem for the {\it nonnegative\/} weight
$|\Phi|^2$.
\vskip13pt 

The  proof of Theorem I makes use of the connection mentioned above 
between the 
boundedness problem for 
the relativistic operator $$\mathcal{H} = \sqrt{-\Delta} + Q: \, W_2^{1/2}
(\R^n) \to W_2^{-1/2}(\R^n),$$
and  its nonrelativistic counterpart, 
$$H = -\Delta + \widetilde Q: \, W_2^1(\R^{n+1}) \to W_2^{-1}(\R^{n+1}).$$
The latter is acting on a pair of  Sobolev spaces of integer order in the
higher dimensional Euclidean space,  and the corresponding potential
$\widetilde Q \in \D'(\R^{n+1})$.  We also employ extensively  
a calculus of 
singular integral,  maximal, and Fourier multiplier operators on the space 
of functions $f \in L_{2, \, loc}(\R^n)$ such that 
$$\int_{\R^n} |f(x)|^2 \, |u(x)|^2 \, dx \le \text{const} \, 
 ||u||^2_{W^{m}_2}, \qquad 
u \in \D(\R^n),$$
developed in \cite{MV1}, \cite{MV2}, and based on the theory 
of Muckenhoupt weights and use of equilibrium measures associated 
with arbitrary compact sets of positive capacity.

Combining  Theorem I with  the characterizations of the inequality
(\ref{E:1.4})  for nonnegative weights established 
earlier (see, e.g., \cite{ChWW}, \cite{Fef}, \cite{KeS},  
\cite{M1}, \cite{M2},
\cite{MV1},  \cite{MV2}, \cite{Ver}) we obtain more explicit 
characterizations
of admissible weights $Q$ stated in Theorem II. \vskip13pt

We now recall  the well-known
isoperimetric inequalities (see, e.g.,  \cite{MSh},  Sec. 2.1.2): 
\begin{align}
 & \text{\rm{cap}} \, (e,  \,
W^{1/2}_2 (\R^n))  \ge c \, |e|^{(n-1)/n}, \quad & \text{diam} \,
(e) \le 1, \qquad &  n \ge 2, \notag \\
 & \text{\rm{cap}} \, (e,  \,
W^{1/2}_2 (\R^1))  \ge \frac {c} {\log \frac 2{|e|}}, 
\quad  & \text{diam} \,
(e) \le 1, \qquad & n =1, \notag
\end{align}
where $|e|$ is Lebesgue measure of a compact set $e \subset \R^n$.
Note that the one-dimensional case 
is special in this setting, since $m = 1/2$ is the critical Sobolev 
exponent 
for $W^{m}_2 (\R^n)$ if $n=1$. Thus, it requires certain modifications in
comparison to  the general case $n \ge 2$.

These estimates together with statement (i) of Theorem II (note that it is
enough  to verify (\ref{E:1.6}) only for compact sets $e$ such that
$\text{diam} \, (e) \le 1$),  yield   sharp sufficient conditions 
for (\ref{E:1.4}) to hold.

\vskip13pt

\noindent {\bf Corollary 1.} 
{\it Suppose $Q \in \D'(\R^n)$, $n\ge 1$. Then 
 $\mathcal{H} = 
\sqrt{- \Delta} + Q$ 
 is  a bounded operator
from  $W^{1/2}_2 (\R^n)$ to $W^{-1/2}_2 (\R^n)$ if  one of the 
following conditions holds:
\begin{equation}\label{E:1.9}
\int_e |\Phi (x)|^2 \, dx \le
c   \, |e|^{(n-1)/n},  \quad  \text{\rm{diam}} \, (e) \le 1, \qquad n
\ge 2, \end{equation}
or
\begin{equation}\tag{\ref{E:1.9}$'$}
\int_e |\Phi (x)|^2 \, dx \le
\frac {c} {\log \frac {2}{|e|}}, \quad  \text{\rm{diam}} \,
(e) \le 1,\qquad n =1,
\end{equation}
where the constant $c$ does not depend on $e \subset \R^n$.
}

\vskip13pt

\noindent {\bf Remark 1.} 
We observe that  (\ref{E:1.9}) holds  if $\Phi \in L_{2n,
\infty} (\R^n) + L_{\infty} (\R^n)$, $n \ge 2$,
where $L_{p, \infty}$ 
denotes the weak $L_p$ (Lorentz) space. Similarly, 
in the one-dimensional case, 
(\ref{E:1.9}$'$) holds if
$\Phi \in L_{1+\epsilon}(\R^1) + L_\infty(\R^1)$, $\epsilon >0$.
\vskip13pt

\noindent {\bf Remark 2.} 
 The class of admissible potentials $Q$ 
satisfying (\ref{E:1.9}) is substantially broader
 than the standard  (in the relativistic case) class 
$Q \in L_n (\R^n) +L_\infty (\R^n)$, $n \ge 2$. 
In particular, it contains highly oscillating  functions 
with significant growth of $|Q|$ at infinity, along with singular
measures and distributions. Similarly, 
in the one-dimensional case, the class of potentials defined by 
(\ref{E:1.9}$'$) is much wider than the standard  class 
$Q \in L_{1+\epsilon}(\R^1) + L_\infty(\R^1)$, $\epsilon >0$. 
 (See \cite{LL}, Sec. 11.3.)
\vskip13pt

These relations, along with sharper estimates  in terms of 
Morrey spaces of negative order  which follow from Theorems I and II, are
discussed in Sec. 3.  They extend significantly  relativistic analogues of 
the Fefferman-Phong class introduced in \cite{Fef}, 
as well as other known classes of admissible potentials.

\vskip13pt
\section{The form boundedness  criterion}
\vskip13pt

For positive integers $m$, the Sobolev space $W_2^m(\R^n)$ is 
defined as  the space
 of weakly differentiable 
 functions  such that 
\begin{equation}\label{E:2.1}
||f||_{W_2^m} = \left [ \int_{\R}  
( |f (x) |^2 + |\nabla^m f(x)|^2 ) \, dx
 \right ]^{\frac 1 2} < \infty.
\end{equation}

More generally, for  real $m>0$,  $W_2^m(\R^n)$ is the space of 
all $f \in L_2(\R^n)$ which can be represented in the form 
$f = (-\Delta + 1)^{-m/2} g$, where $g \in L_2(\R^n)$. Here 
$(-\Delta + 1)^{-m/2} g = J_m \star g$ is the convolution of $g$ with 
the Bessel kernel $J_m$ of order $m$, and 
$||f||_{W_2^m} = ||g||_{L_2^m}$ 
 (see  \cite{M2}, \cite{St1}). This definition 
is consistent with the previous one for  integer $m$, and
defines an equivalent norm on $W_2^m(\R^n)$. 
Note that another equivalent norm on $W_2^m(\R^n)$ is 
given by 
$$||f||_{W_2^m} = ||f||_{L_2} + \left \Vert \, |D|^m f \, 
\right \Vert_{L_2},  
\quad f \in W_2^m(\R^n),$$
where $|D| = (-\Delta)^{1/2}$.

The dual space $W_2^{-m} (\R^n) = W_2^{m}(\R^n)^*$ can be identified 
with  the
space of  distributions $f$ of the form $f = (-\Delta + 1)^{m/2} g$, 
where $g \in L_2(\R^n)$.

Let $\gamma \in \D'(\R^n)$ be a (complex-valued) distribution on $\R^n$. 
We will use the same notation for  the  corresponding multiplication 
operator 
$\gamma: \, \D(\R^n) \to \D'(\R^n)$ defined by:
$$\langle \gamma u, v\rangle  = \langle  \gamma, \bar u \, v \rangle 
\qquad u, v \in \D(\R^n).$$

For $m, \, l \in \R$, we 
denote by $\text{Mult} \, (W_2^m \to W_2^l)$ the class of
bounded  multiplication operators (multipliers)
 from $W_2^m$ to $W_2^l$ generated by 
$\gamma \in
\D'(\R^n)$   so that  the corresponding  sesquilinear form 
$\langle \gamma \,  \cdot, \, \cdot \rangle$ is bounded:
\begin{equation}\label{E:2.2}
 |\langle \gamma \, u, \, v \rangle | = |\langle \gamma, \, \bar u \, v
\rangle | \leq C  \, ||u||_{W_2^{m}} \, ||v||_{W_2^{-l}}, 
\qquad u, v \in \D(\R^n),
\end{equation}
where $C$ does not depend on $u, v$. The {\it multiplier norm}  
denoted by  $||\gamma||_{W_2^m \to W_2^l}$ 
is equal to the 
least bound $C$ 
 in the preceding inequality. 

It is easy to see that, in the case $l=-m$, (\ref {E:2.2}) 
is equivalent to the
quadratic  form inequality:
\begin{equation}\tag{\ref{E:2.2}$'$}
|\langle \gamma \, u, \,u \rangle | = |\langle \gamma, \, |u|^2
\rangle | \le C'  \, ||u||^2_{W_2^{m}}, \qquad u \in
\D(\R^n). 
\end{equation}
To verify this, suppose that 
$||u||_{W_2^{m}} \le 1$, 
$||v||_{W_2^{m}} \le 1$, where  $u, \, v \in \D(\R^n)$. Applying
(\ref{E:2.2}$'$) together with  the polarization identity:
$$
\bar u \, v = \frac 1 4 \left( \, |u+v|^2 - |u-v|^2 -i |u - iv|^2 + i |u+iv|^2
\,\right), $$
and the parallelogram identity, we get:
\begin{align}
|\langle \gamma, \, \bar u \, v\rangle | 
& \le  \frac {C'} 4 \left (||u+v||^2_{W_2^{m}}
+ ||u -v||^2_{W_2^{m}} + ||u + i v||^2_{W_2^{m}} + 
||u - i v ||^2_{W_2^{m}}
\right)\notag\\ & \le 2 C'. 
\notag
\end{align} 
Hence, (\ref{E:2.2}) holds for $l=-m$ with $C = 2 C'$. Moreover, the least
bound $C'$ in  (\ref{E:2.2}$'$) satisfies the inequality: 
$$C' \le ||\gamma||_{W_2^m \to
W_2^{-m}} \le 2 C'.$$

Let $|D| = (-\Delta)^{1/2}$.   
We define the relativistic Schr\"odinger operator as 
$$\mathcal{H} = |D| + Q : \, \D(\R^n) \to \D'(\R^n),$$
(see \cite{LL}), where $Q : \, \D(\R^n) \to\in \D'(\R^n)$ is  
a multiplication 
operator defined by $Q \in \D'(\R^n)$.  
It is well-known that actually 
$|D|$ is a bounded operator from $W_2^{1/2} (\R^n)$ to
$W_2^{-1/2}(\R^n)$. Thus,  $\mathcal{H}$ can be extended
to a bounded operator: 
$$\mathcal{H} : \, W_2^{1/2} (\R^n) \to W_2^{-1/2}(\R^n),$$ 
if and only if 
$Q \in \text{Mult} \, (W_2^{1/2}(\R^n) \to W_2^{-1/2}(\R^n))$, or, 
equivalently, if the quadratic form inequality (\ref{E:2.2}$'$)
holds for $\gamma =Q$ and $m=1/2$. 

From the preceding discussion it follows that  
$\mathcal{H} : \, W_2^{1/2} (\R^n) \to W_2^{-1/2}(\R^n)$ 
is bounded if and only if 
\begin{equation}\label{E:2.3}
|\langle Q \, u, \,u \rangle |
 \le a \,  \langle |D| \, u, \, u\rangle  + 
b \,  \langle  u, \,u \rangle,
\qquad u \in \D(\R^n), 
\end{equation}
for some $a, \, b>0$. By definition this means that  $Q$ is
{\it relatively  form bounded\/} with respect to $|D|$. 

In particular, 
if $Q$ is real-valued, and  $0<a<1$ in the preceding inequality, 
then by
the so-called  KLMN Theorem (\cite{RS}, Theorem X.17), 
$\mathcal{H} = |D| + Q$ is defined as a unique self-adjoint 
operator  such that 
$$\langle \mathcal{H} u, \, v\rangle = 
\langle |D| \, u, \, v\rangle + \langle Q \, u, v\rangle, \qquad 
u \in \D(\R^n).
$$
For complex-valued $Q$ such that (\ref{E:2.3}) holds with $0<a<1/2$, 
it follows that $\mathcal{H} = |D| + Q$,  understood in a similar sense,  
 is  an
$m$-sectorial operator (\cite{EE}, Theorem IV.4.2).

In the case where $Q \in L_{1, \, loc}(\R^n)$, 
 (\ref{E:2.3}) is equivalent 
to 
the  
inequality: 
\begin{equation}\label{E:2.4}
\left \vert \int_{\R^n} |u(x)|^2 \, Q(x) \, dx\right 
\vert \leq \text{\rm{const}} \, 
||u||^2_{W_2^{1/2}}, \quad u \in \D(\R^n), 
\end{equation} 
and hence  to the boundedness 
of the corresponding sesquilinear form:
$$
 \left \vert \int_{\R^n} 
  u(x)  \, \overline {v(x)} \, \, Q(x) \, dx\right 
\vert \leq \text{\rm{const}} 
\, ||u||_{W_2^{1/2}(\R^n)} \, ||v||_{W_2^{1/2}(\R^n)}, 
$$
where the constant is independent of $u, v \in \D(\R^n)$.

Our characterization of potentials $Q$ such that 
$\mathcal{H}: \, W_2^{1/2}(\R^n) \to W_2^{-1/2}(\R^n)$ is based on 
a series 
of lemmas and propositions presented below, and the results of 
\cite{MV2} for the nonrelativistic Schr\"odinger operator.

By $L_{2, unif} (\R^n)$, we denote the class of $f \in L_{2, loc} (\R^n)$ 
such that
\begin{equation}\label{E:2.5}
||f||_{L_{2, unif}} = \sup_{x \in \R^n} \, 
||\chi_{B_1 (x)} \, f||_{L_2(\R^n)} < \infty,
\end{equation} 
where $B_r(x)$ denotes a Euclidean ball of radius $r$ centered at $x$.

\vskip13pt

 \noindent {\bf Lemma 2.1.}
 {\it  
 Let $0<l<1$, and $m>l$.  Then 
$\gamma \in \text{\rm{Mult}} \, (W_2^m \to W_2^l)$ 
if and only if $\gamma \in W_2^m \to W_2^{m-l}$, and $|D|^l \gamma 
\in \text{\rm{Mult}} \, (W_2^{m} \to L_2)$.
  Moreover, 
  \begin{equation}\label{E:2.6}
  \left \Vert \gamma\right \Vert_{W_2^m \to W_2^l} \thicksim 
\left \Vert |D|^l \gamma \right \Vert_{W_2^{m} \to L_2} + 
\left \Vert \gamma\right \Vert_{W_2^m \to W_2^{m-l}}.   \end{equation}
  \vskip13pt

 \begin{proof} We first prove the lower estimate for 
$\left \Vert\gamma\right \Vert_{W_2^m \to W_2^l}$:
\begin{equation}\label{E:2.7}
\left \Vert |D|^l \gamma \right \Vert_{W_2^{m} \to L_2} + 
 \left \Vert \gamma\right \Vert_{W_2^m \to W_2^{m-l}} \le c \, 
\left \Vert\gamma\right \Vert_{W_2^m \to W_2^l}. 
\end{equation}
Here and below $c$ denotes a constant which 
depends only on $l, m$, and
$n$.  

Let $u \in C^\infty_0(\R^n)$. 
Using the integral representation (which follows by inspecting 
the Fourier  transforms of both sides),
\begin{equation}\label{E:2.8}
|D|^l u (x) =
 c (n,l) \, \int_{\R^n} \frac{u(x) - u(y)} {|x-y|^{n+l}} dy,
\end{equation}
we obtain:
\begin{align}
& |D|^l \, (\gamma \, u) (x) - \gamma(x) \, |D|^l \, u (x) - u(x) \, 
|D|^l \, \gamma (x)  \notag \\ & = - c (n, l) \, 
\int_{\R^n} \frac{(u(x) - u(y))(\gamma (x) - \gamma 
(y))} {|x-y|^{n+l}} dy. \notag
\end{align}
Hence,
\begin{equation}\label{E:2.9}
\left \vert \, |D|^l \, (\gamma \, u) - \gamma \, |D|^l \, 
u  - u \, 
|D|^l \gamma  \right \vert \le c \, \D_{l/2} \, u \cdot 
\D_{l/2} \, \gamma, 
\end{equation}
where
$$
\D_{s} \, u (x) = \left ( \int_{\R^n}
 \frac{|u(x) - u(y)|^2} {|x-y|^{n+2s}} dy  
\right)^{\frac 1 2}, \quad s>0.
 $$
Next, we estimate:
\begin{align}
& ||u \cdot |D|^l  \gamma ||_{L_2}  \le 
\left \Vert \, |D|^l  (\gamma \, u) \right \Vert_{L_2} +  \left \Vert 
\gamma \, |D|^l  u \right \Vert_{L_2} + 
c \, ||\D_{l/2}  u  \cdot  \D_{l/2}  \gamma ||_{L_2} \notag \\
& \le \, ||\gamma u||_{W_2^l} + ||\gamma||_{W_2^{m-l} \to L_2} \,
\left \Vert \, |D|^l  u \right \Vert_{W_2^{m-l}} + 
c \, ||\D_{l/2}  u  \cdot  \D_{l/2}  \gamma ||_{L_2} \notag \\
& \le \, ||\gamma||_{W_2^m \to W_2^l} \, ||u||_{W_2^m} +
||\gamma||_{W_2^{m-l} \to L_2} \, ||u||_{W_2^m} +  c \, ||\D_{l/2}  u 
\cdot  \D_{l/2}  \gamma ||_{L_2}\notag \\
& \le c \, ||\gamma||_{W_2^m \to W_2^l} \, ||u||_{W_2^m}  + 
 c \, ||\D_{l/2}  u \cdot  \D_{l/2}  \gamma ||_{L_2}.\notag
\end{align}
In the last line we have used the known inequality (\cite{MSh}, Sec. 2.2.2):
$$||\gamma||_{W_2^{m-l} \to L_2}  \le c \, ||\gamma||_{W_2^m \to W_2^l}.$$

To estimate the  term $||\D_{l/2}  u \cdot  \D_{l/2}  \gamma ||_{L_2}$,
we apply  the pointwise estimate (Lemma 1 in \cite{MSh}, Sec. 3.1.1):
$$\D_{l/2}  u \le  \, J_{s} \D_{l/2} ( (-\Delta +1)^{s/2}\, u),$$
with $s=m - l/2$, where 
$J_s = (-\Delta +1)^{-s/2}$ is the Bessel potential of order 
$s$. Hence
\begin{align}
& ||\D_{l/2}  u \cdot  \D_{l/2}  \gamma ||_{L_2}  \le  
\, ||J_{m-l/2} \D_{l/2} ( (-\Delta +1)^{m/2 -l/4}\, u)) \cdot 
\D_{l/2}  \gamma ||_{L_2} \notag\\
& \le \, c \, || \D_{l/2} \gamma ||_{W_2^{m-l/2} \to L_2} \, 
||J_{m-l/2} \D_{l/2} ( (-\Delta +1)^{m/2 -l/4}\, u))||_{W_2^{m-l/2}} 
\notag\\
& \le \, c \, || \D_{l/2} \gamma ||_{W_2^{m-l/2} \to L_2} \, 
|| \D_{l/2} (-\Delta +1)^{m/2 -l/4} u||_{L_2}  \notag\\
& \le \, c \, || \D_{l/2} \gamma ||_{W_2^{m-l/2} \to L_2} \, 
||u||_{W_2^m}.\notag
\end{align}
We next  show that
$$|| \D_{l/2} \gamma ||_{W_2^{m-l/2} \to L_2}  \le c \, 
||\gamma||_{W_2^{m} \to W_2^{l}}.$$
By the Lemma  in \cite{MSh}, Sec. 3.2.5 in the case $p=2$, we have:
$$|| \D_{l} \gamma ||_{W_2^{m} \to L_2} + ||\gamma||_{W_2^{m-l} \to
L_2} \le c \, 
||\gamma||_{W_2^{m} \to W_2^{l}},$$
where $m \ge l>0$.
Applying the preceding estimate 
 with $m-l/2$ in place of
$m$  and $l/2$ in place of $l$ respectively, we get:
$$|| \D_{l/2} \gamma ||_{W_2^{m-l/2} \to L_2} + ||\gamma||_{W_2^{m-l} \to
L_2} \le c \, 
||\gamma||_{W_2^{m-l/2} \to W_2^{l/2}}.$$
Now by interpolation, 
$$||\gamma ||_{W_2^{m-l/2} \to W_2^{l/2}} 
\le ||\gamma||^{1/2}_{W_2^{m-l} \to L_2} \, ||\gamma||^{1/2}_{W_2^{m} \to
W_2^l}.$$ Since $||\gamma||_{W_2^{m-l} \to
L_2}\le c \, ||\gamma||_{W_2^{m} \to
W_2^{l}}$, it follows that 
$$||\gamma ||_{W_2^{m-l/2} \to W_2^{l/2}} \le c \, ||\gamma||_{W_2^{m} \to
W_2^{l}}.
$$
Hence, 
$$|| \D_{l/2} \gamma ||_{W_2^{m-l/2} \to L_2}  \le c \, ||\gamma
||_{W_2^{m-l/2} \to W_2^{l/2}} \le c \, ||\gamma||_{W_2^{m} \to
W_2^{l}}.$$
Combining these estimates, we obtain:
$$\Vert u \cdot |D^l| \gamma\Vert_{L_2} \le c \,
||\gamma||_{W_2^{m} \to W_2^{l}} \, ||u||_{W_2^m},$$
which is equivalent to the inequality  
$$\Vert |D^l| \gamma \Vert_{W_2^{m} \to L_2} \le c \, 
||\gamma||_{W_2^{m} \to W_2^{l}}.$$
 This, together with the 
inequality 
$||\gamma||_{W_2^{m-l} \to
L_2} \le c \, 
||\gamma||_{W_2^{m} \to W_2^{l}}$
used above, completes the proof of  (\ref{E:2.7}). 

We now prove the upper estimate
\begin{equation}\label{E:2.10}
  \left \Vert \gamma\right \Vert_{W_2^m \to W_2^l} \le
c \, \left (\left \Vert |D|^l \gamma \right \Vert_{W_2^{m} \to L_2} + 
||\gamma||_{W_2^{m-l} \to
L_2}  \right).  
 \end{equation}
By (\ref{E:2.9}), 
$$
\left \Vert  |D|^l  (\gamma u) \right\Vert_{L_2} 
 \le \left \Vert \gamma |D|^l  u\right
\Vert_{L_2}   + \left \Vert |D|^l \gamma \cdot u \right \Vert_{L_2}
+ c \, ||\D_{l/2} u \cdot \D_{l/2} \gamma||_{L_2}.
$$
Using an elementary estimate $||u||_{W_2^{m-l}} \le c \, ||u||_{W_2^{m}}$, 
we have: 
$$\left \Vert \gamma u \right
\Vert_{L_2} \le ||\gamma||_{W_2^{m-l}\to L_2} \,
||u||_{W_2^{m-l}} \le c \, ||\gamma||_{W_2^{m-l}\to L_2} \,
||u||_{W_2^{m}}.$$
From these inequalities, combined with
the estimate  
$$||\D_{l/2} u \cdot \D_{l/2} \gamma||_{L_2} \le 
c \,||\gamma||_{W_2^{m-l/2}\to W_2^{l/2}} \,
||u||_{W_2^m}
$$ 
established above, 
it follows:
\begin{align}
  ||\gamma u||_{W_2^l} \le  \, 
& c \, (||\gamma||_{W_2^{m-l}\to L_2} \, ||u||_{W_2^m}  + 
\left \Vert |D|^l \gamma \right \Vert_{W_2^m \to L_2} 
\, ||u||_{W_2^m}) \notag  \\
 + & c \, ||\gamma||_{W_2^{m-l/2}\to W_2^{l/2}} \, ||u||_{W_2^m}.\notag
\end{align}
As above, by an interpolation argument,
$$||\gamma||_{W_2^{m-l/2}\to W_2^{l/2}} \le 
||\gamma||^{1/2}_{W_2^{m-l}\to L_2} \, ||\gamma||^{1/2}_{W_2^{m}\to
W_2^{l}}.$$ 
Thus,
$$
 \left \Vert \gamma\right \Vert_{W_2^m \to W_2^l} \le
c \left (\left \Vert |D|^l \gamma \right \Vert_{W_2^{m} \to L_2} + 
||\gamma||_{W_2^{m-l}\to L_2}  + ||\gamma||^{1/2}_{W_2^{m-l}\to L_2} \,
||\gamma||^{1/2}_{W_2^{m}\to W_2^{l}} \right). 
$$
Clearly, the preceding estimate yields:
$$
\left \Vert \gamma\right \Vert_{W_2^m \to W_2^l} \le
c \left (\left \Vert |D|^l \gamma \right \Vert_{W_2^{m} \to L_2} + 
 ||\gamma||_{W_2^{m-l}\to L_2}\right).
$$
This completes the proof of Lemma 2.1. 
\end{proof}

\vskip13pt

\noindent {\bf Lemma 2.2.}
 {\it Let $0<l<1$, and $m>l$.  Then 
$\gamma \in \text{\rm{Mult}} \, (W_2^m \to W_2^l)$ 
if and only if $(-\Delta + 1)^{l/2} \gamma \in  \text{\rm{Mult}} \, 
(W_2^{m} \to L_2)$, and 
} 
\begin{equation}\label{E:2.11}
  ||\gamma||_{W_2^m \to W_2^l} \thicksim 
||(-\Delta + 1)^{l/2} \gamma ||_{W_2^{m} \to L_2}.
  \end{equation}
  \vskip13pt

 \begin{proof} We denote by $M$ the Hardy-Littlewood maximal operator:
$$M f(x) = \sup_{r>0}  \frac {1} {|B_r(x)|} \int_{B_r(x)} |f(y)| \, dy, 
\qquad x \in
\R^n.$$ 
Recall that 
a nonnegative weight $w \in L_{1, \,loc} (\R^n)$ is said to be in the 
Muckenhoupt class $A_1(\R^n)$ if 
$$M w(x) \le \text{const} \, w(x) \qquad \quad \text{a.e.}$$ 
The least  constant on the right-hand side of the 
preceding inequality is called the $A_1$-bound of 
$w$.

We will need the following 
statement 
established earlier   in \cite{MV1},   Lemma 3.1.
\vskip13pt

\noindent {\bf Lemma 2.3.}  {\it Let
$\gamma \in \text{\rm{Mult}} \, (W_p^m \to L_p)$, where $1<p<\infty$, 
and $m > 0$. Suppose that
$T$ is  a bounded operator on the weighted space $L_p (w)$ for every 
$w \in A_1(\R^n)$.  Suppose additionally that, for all $f \in L_p(w)$,
the inequality $$|| T f||_{L_p(w)} \le C \, ||f||_{L_p(w)} $$
holds with the constant $C$ which depends only on the $A_1$-bound of the 
weight $w$. Then $T \gamma \in \text{\rm{Mult}} \, (W_p^m \to L_p)$, and
$$
||T \gamma||_{W_p^{m} \to L_p} \le C_1 ||\gamma||_{W_p^{m} \to L_p},
$$
where the constant $C_1$ does not depend on 
$\gamma$.}
  \vskip13pt
  
We will also need a Fourier multiplier theorem  of Mikhlin type for 
$L_p$ spaces with weights. Let $m \in L_\infty(\R^n)$. Then the Fourier 
multiplier operator with symbol $m$ is defined on $L_2(\R^n)$ 
 by $T_m  = \mathcal{F}^{-1}
\,   m \, \mathcal{F}$, where 
$\mathcal{F}$ and $\mathcal{F}^{-1}$ are respectively the direct and 
inverse Fourier transforms. 

The following lemma follows from the results of Kurtz and Wheeden 
 \cite {KWh}, Theorem 1.

\vskip13pt
\noindent {\bf Lemma 2.4.}  {\it Suppose $1<p< \infty$ and $w \in 
A_1(\R^n)$. Suppose that $m \in C^\infty(\R^n\setminus\{0\})$ 
 satisfies the 
Mikhlin multiplier condition:
\begin{equation}\label{E:2.12}
|D^\alpha \, m (x)| 
\le C_\alpha \, |x|^{-|\alpha|}, \qquad x \in
\R^n\setminus\{0\},
 \end{equation}
 for every multi-index $\alpha$ such that $0 \le |\alpha| \le n$. 
 Then  
 the inequality  
$$ ||T_m \, f ||_{L_p(w)} \le C \, ||f ||_{L_p(w)}, 
\qquad f \in L_p (w) \cap L_2(\R^n),$$
holds with the constant that depends only on $p$, $n$, 
the $A_1$-bound of
$w$,  and the constant $C_\alpha$ in \rm{(\ref{E:2.12})}.} 
\vskip13pt

\noindent {\bf Corollary 2.5.}  {\it Suppose $1<p< \infty$ and $w \in 
A_1(\R^n)$. Suppose $0<l \le 2$. Define 
\begin{equation}\label{E:2.13}
m_l (x) = (1 + |x|^2)^{l/2} - |x|^l.
\end{equation}
Then 
\begin{equation}\label{E:2.14}
||T_{m_l}\, f||_{L_p(w)} \le C \, ||f||_{L_p(w)}, \qquad
 f \in L_p (w) \cap L_2(\R^n),
\end{equation}
where the constant $C$ depends only on $l$, $p$, $n$, and 
the $A_1$-constant of $w$.}
\vskip13pt
\noindent {\bf Remark.} It is well known that in the unweighted case the
operator $T_{n_l} = (1-\Delta)^{-l/2}\, T_{m_l}$,   is bounded on 
$L_p(\R^n)$
for all  $l >0$ and $1 \le p \le \infty$,  including the endpoints
(\cite{St1},  Sec. 5.3.2, Lemma 2). 
\vskip13pt      
\noindent{\it Proof of Corollary 2.5.}  Clearly, 
$$0 \le m_l(x) \le C \, (1 + |x|)^{l-2}, \qquad x \in \R^n.$$ 
Furthermore, it is easy to see by induction that, for any multi-index 
$\alpha$, $|\alpha| \ge 1$, we have the following estimates:
$$|D^\alpha \, m_l (x)| 
\le C_{\alpha, l} \, |x|^{l - 2 - |\alpha|}, 
\qquad |x| \to \infty,$$
and 
$$|D^\alpha \, m_l (x)| 
\le C_{\alpha, l} \, |x|^{l  - |\alpha|}, 
\qquad |x| \to 0.$$
Since $0<l \le 2$, from this it follows that $m_l$ satisfies 
(\ref{E:2.12}), and hence by Lemma 2.4 the inequality 
$$
||T_{m_l}\, f||_{L_p(w)} \le C \, ||f||_{L_p(w)}
$$ 
holds with a constant that depends only on $l$, $p$, and 
the $A_1$-bound of $w$. 
\qed
\vskip13pt
We are now in a position to complete the proof of Lemma 2.2.
Suppose that $\gamma \in \text{\rm{Mult}} \, (W_2^m \to W_2^l)$, where 
$m>1$ and 
$0<l<1$. By Corollary 2.5, the operator $T_{m_l} = 
(1-\Delta)^{l/2} - |D|^l$
is bounded on  $L_2(w)$ for every $w \in A_1$, and its norm is bounded 
by a constant which depends only on $l$, $n$, and the $A_1$-bound of $w$. 
Hence by 
Lemma 2.3 it follows that 
$\left ( (1-\Delta)^{l/2} - |D|^l\right) \gamma \in \text{\rm{Mult}} 
\, (W_2^m \to L_2)$, and 
$$
\Vert \left ( (1-\Delta)^{l/2} - |D|^l\right)  \gamma\Vert_{W_2^{m} 
\to L_2} \le c \,
||\gamma||_{W_2^{m} \to L_2}, $$
where $c$ depends only on $l$, $m$, and $n$. 

Clearly, 
$||\gamma||_{W_2^{m} \to L_2} \le ||\gamma||_{W_2^{m} \to W_2^l}.$
Using these estimates and   Lemma 2.1, we obtain:
$$
||(1-\Delta)^{l/2} \, \gamma||_{W_2^{m} \to L_2} \le c \, 
\left ( \Vert |D|^l \gamma \Vert_{W_2^{m} \to L_2} +  
||\gamma||_{W_2^{m} \to L_2}  
\right) 
\le c \, ||\gamma||_{W_2^{m} \to W_2^l}.
$$

Conversely, suppose that $(1-\Delta)^{l/2} \, \gamma \in \text{\rm{Mult}} 
\, (W_2^m \to L_2)$. It follows from the above estimate of 
$\Vert \left ( (1-\Delta)^{l/2} - |D|^l\right)  
\gamma \Vert_{W_2^{m} \to L_2}$
 that 
$$\left \Vert |D|^{l} \gamma| \right \Vert_{W_2^{m} \to L_2} \le c \, 
\left ( ||(1-\Delta)^{l/2} \, \gamma ||_{W_2^{m} \to L_2} + 
||\gamma||_{W_2^{m} \to L_2} \right).$$
Obviously, 
$||\gamma||_{W_2^{m} \to L_2} \le c \, ||\gamma||_{W_2^{m-l} \to L_2}.$ 
Applying again  Lemma 2.1 together with the preceding estimates, we have:
\begin{align}
 ||\gamma||_{W_2^{m} \to W_2^l} & \le c \,
 \left ( \left \Vert \, |D|^{l} \gamma| \right \Vert_{W_2^{m} \to L_2}
+ ||\gamma||_{W_2^{m-l} \to L_2} \right) \notag \\ & \le 
c \, \left ( ||(1-\Delta)^{l/2} \, \gamma ||_{W_2^{m} \to L_2} + 
||\gamma||_{W_2^{m-l} \to L_2} \right).\notag
\end{align}

It remains to obtain the estimate 
$$||\gamma||_{W_2^{m-l} \to L_2} \le c \,
 ||(1-\Delta)^{l/2} \, \gamma ||_{W_2^{m} \to L_2}, 
$$
whose proof is similar to the argument used in \cite{MSh}, 
Sec. 2.6, and is outlined below. 

Since $(1-\Delta)^{l/2} \, \gamma \in \text{\rm{Mult}} 
\, (W_2^m \to L_2)$, it follows that 
$$\int_{e} |(1-\Delta)^{l/2} \, \gamma|^2 \, dx \le \,
 ||(1-\Delta)^{l/2} \, \gamma||_{W_2^{m} \to L_2}^2 \, 
\text{cap} \, (e, \, W_2^m),
$$
for every compact set $e \subset \R^n$. 
Hence, for every ball $B_r(a)$, 
$$\int_{B_r(a)} |(1-\Delta)^{l/2} \, \gamma|^2 \, dx \le c \, 
||(1-\Delta)^{l/2} \, \gamma||_{W_2^{m} \to L_2}^2 \, r^{n-2m}, 
\quad 0<r \le 1,
$$
and  in particular 
$$
||(1-\Delta)^{l/2} \, \gamma||_{L_2, unif} \le c \, 
||(1-\Delta)^{l/2} \, \gamma ||_{W_2^{m} \to L_2}.$$

Notice that $ \gamma =  J_l \, (1-\Delta)^{l/2} \, \gamma$, 
where the Bessel potential $J_l = (1-\Delta)^{-l/2}$ can be 
represented as a convolution 
operator, $J_l f = G_l \star f$. Here $G_l$ is a positive 
radially decreasing function whose behavior at $0$ and infinity 
respectively is given 
by 
\begin{align}
G_l(x)& \asymp |x|^{l-n} \quad  \text{as} \quad x\to 0,
 \qquad \text{if} \quad
0<l<n, \notag \\
G_l(x) & \asymp |x|^{(l-n-1)/2} \, e^{-|x|} \quad  \text{as} 
\quad |x|\to +\infty.\notag
\end{align}
From this, it is easy to derive the  pointwise estimate
\begin{align}
|\gamma (x)| & \le  \int_{\R^n} G_l (x-t) \, 
 |(1-\Delta)^{l/2} \, \gamma (t)| \, dt 
\notag \\ & 
\le c \, \left (\int_{|z|\le 1} \frac {|(1-\Delta)^{l/2} \, \gamma(x+z)|}
{|z|^{n-l}} \, dz + ||(1-\Delta)^{l/2} \, \gamma||_{L_2, unif} \right).
\notag 
\end{align}
Using Hedberg's inequality together 
with the preceding pointwise estimate, 
as in the proof of Lemma 2.6.2 in \cite{MSh}, 
we deduce:
\begin{align}
& |\gamma (x)| \le c \, ( M \, (1-\Delta)^{l/2} \, \gamma (x))^{1- \frac l m}  
\left (  \sup_{0<r\le 1, \, a \in \R^n} \frac 
{\int_{B_r(a)} |(1-\Delta)^{l/2} \, \gamma|^2 \, dy} {r^{n-2m}}   
\right)^{\frac l {2m}} \notag \\
&+ c \, ||(1-\Delta)^{l/2}  \gamma||_{L_2, unif}  
\le c \, 
( M \, (1-\Delta)^{l/2} \, \gamma (x))^{1- \frac l m}  
||(1-\Delta)^{l/2} \gamma||_{W_2^{m} \to L_2}^{\frac l m} \notag \\
&+ c \, ||(1-\Delta)^{l/2}  \gamma||_{W_2^{m} \to L_2}, \notag 
\end{align}
where $M$ is the Hardy-Littlewood maximal operator. 
Using the preceding estimates, together with 
the boundedness of 
 $M$ on the space $\text{\rm{Mult}} \, (W_2^m \to L_2)$ (see details 
in \cite{MSh}, Sec. 2.6) we obtain:
$$\Vert |\gamma|^{\frac m {m-l}} \Vert_{W_2^m \to L_2}^{1- \frac l m} 
\, \le c \, 
 ||(1-\Delta)^{l/2} \,\gamma||_{W_2^{m} \to L_2}.$$
By Lemma 2 in   \cite{MSh}, Sec. 2.2.1, it follows: 
$$\Vert \gamma \Vert_{W_2^{m-l} \to L_2} \le c \, 
\Vert |\gamma|^{\frac m {m-l}} \Vert_{W_2^m \to L_2}^{1- \frac l m} 
\, \le c \, 
 ||(1-\Delta)^{l/2} \,\gamma||_{W_2^{m} \to L_2}.$$
The proof of Lemma 2.2 is complete. 
\end{proof}
\vskip13pt

\noindent {\bf Theorem 2.6.}
 {\it 
Let $\gamma \in \D'(\R^n)$. Then  
$\gamma \in \text{\rm{Mult}} \, (W_2^{1/2} (\R^n)\to W_2^{-1/2}(\R^n))$ 
if and only if $\Phi = (-\Delta +1)^{-1/4} \gamma
 \in \text{\rm{Mult}} \, (W_2^{1/2}(\R^n) \to L_2(\R^n))$. Furthermore,
$$
||\gamma||_{W_2^{1/2} \to W_2^{-1/2}} \thicksim 
||\Phi||_{W_2^{1/2} \to L_2}.$$}
  \vskip13pt

 \begin{proof} To prove the ``if'' part, it suffices to verify
that, for every $u \in C^\infty_0 (\R^n)$ and 
 $\Phi = (-\Delta +1)^{-1/4} \gamma
 \in \text{\rm{Mult}} \, (W_2^{1/2} \to L_2)$, the inequality
\begin{equation}\label{E:2.15}
 \left  \vert \int_{\R^n} |u|^2  \gamma \, 
 \right  \vert  \le C \, || \Phi||_{W_2^{1/2} \to L_2} \, 
||u||^2_{W_2^{1/2}}
  \end{equation}
holds. Here the integral on the left-hand side 
is understood  in the sense of 
quadratic forms:
$$  \int_{\R^n} |u|^2  
\gamma = 
\langle  \, \gamma u, \, u \rangle,  
$$
where $\langle  \, \gamma \cdot, \, \cdot \rangle $ is the quadratic 
form associated 
with the multiplier operator $\gamma$, 
as explained in detail in \cite{MV2}.

Since $\gamma = (-\Delta +1)^{1/4} \Phi$, we have:
\begin{align}
\left  \vert \int_{\R^n} |u|^2  \, \gamma 
 \right  \vert
& =  \left  \vert \int_{\R^n} 
(-\Delta +1)^{1/4} \Phi \cdot
\, |u|^2  \right  \vert \notag \\
& \le   \left  \vert \int_{\R^n} 
\left((-\Delta +1)^{1/4} - |D|^{1/2} \right) \Phi \cdot
\, |u|^2 \right \vert  + \left  \vert \int_{\R^n} |D|^{1/2}  \,   \Phi \cdot
\, |u|^2  \right \vert. \notag
\end{align}

Note that $(-\Delta +1)^{1/4} - |D|^{1/2} = T_{m_{1/2}}$, where 
$T_{m_l}$ is the Fourier multiplier operator defined by 
(\ref{E:2.13}). By Corollary 2.5,  $T_{m_{1/2}}$ 
is a bounded operator on $L_2(w)$ for any 
$A_1$-weight $w$, and its norm depends only on the $A_1$-bound 
of $w$. Hence by Lemma 2.3 it follows 
that 
$\left ((-\Delta +1)^{1/4} - |D|^{1/2} \right ) \Phi \in \text{\rm{Mult}} \,
(W_2^{1/2} \to L_2)$,  and 
$$
||\left ((-\Delta +1)^{1/4} - |D|^{1/2}\right) \Phi||_{W_2^{1/2} \to L_2}
 \le C \, || \Phi||_{W_2^{1/2} \to L_2}.
$$ 
Using this estimate and the Cauchy-Schwarz inequality, we get 
\begin{align}
& \left  \vert \int_{\R^n} 
\left ((-\Delta +1)^{1/4}  -  |D|^{1/2} \right) \Phi \cdot
\,  |u|^2  \right  \vert  \notag \\
& \le C \, ||((-\Delta +1)^{1/4} -  |D|^{1/2}) \Phi \cdot u||_{L_2} 
\, ||u||_{L_2} \notag \\ & \le C \,  
|| \Phi||_{W_2^{1/2} \to L_2} ||u||^2_{W_2^{1/2}}.\notag
\end{align}

Hence, in order to prove (\ref{E:2.15}) it suffices 
to establish the
inequality:  
\begin{equation}\label{E:2.16}
  \left  \vert \int_{\R^n} 
|D|^{1/2} \, \Phi \cdot
\, |u|^2 
 \right  \vert\le C \, || \Phi||_{W_2^{1/2} \to L_2} \, 
||u||^2_{W_2^{1/2}}.
  \end{equation}
By duality, 
$$\left  \vert \int_{\R^n} 
|D|^{1/2} \, \Phi \cdot
\, |u|^2 
 \right  \vert = \left  \vert \int_{\R^n} \Phi (x) \, 
(|D|^{1/2} 
\, |u|^2 )(x)
 \, dx 
 \right  \vert,$$
where $\Phi \in L_{2, loc}$, and the integral on the right-hand 
side is well-defined (see details in \cite{MV2}). 

Notice that, for $u \in C^\infty_0 (\R^n)$, 
$$
|D|^{1/2}
\, |u|^2 (x) = c \,  \int_{\R^n}  \frac {|u(x)|^2 - |u(y)|^2}
{|x-y|^{n + 1/2}} \, dy.
$$
Using the identity $|a|^2 -|b|^2 = |a-b|^2 
- 2 \text{Re} \, [\bar b \, (b-a)]$ with $b=u(x)$ and $a=u(y)$, 
and integrating against $\frac { dy}{|x-y|^{n + 1/2}}$,
we get:
\begin{align}
 \int_{\R^n}  \frac {|u(x)|^2 - |u(y)|^2}
{|x-y|^{n + 1/2}} \, dy &= \int_{\R^n}  \frac {|u(x) - u(y)|^2}
{|x-y|^{n + 1/2}} \, dy   \notag \\ 
& -2 \, \text{Re} \left [ \overline{u(x)} 
\int_{\R^n}  \frac {u(x) - u(y)}
{|x-y|^{n + 1/2}} \, dy\right ].\notag
\end{align}
Hence,
\begin{align}
  \left \vert \, |D|^{1/2}
\, |u|^2 (x)\right \vert & \le  
c \, \left ( 2 \, |u(x)| \, \left \vert 
\int_{\R^n}  \frac {u(x) - u(y)}
{|x-y|^{n + 1/2}} \, dy \right \vert 
+  \int_{\R^n}  \frac {|u(x) - u(y)|^2}
{|x-y|^{n + 1/2}} \, dy 
  \right )  \notag  \\
& = 2 c \, |u(x)|\left \vert \, |D|^{1/2}
\, u (x)\right \vert + c \, | \D_{1/4} u(x)|^2.\notag
\end{align}
Using the preceding inequality, we estimate:
\begin{align}
 & \left \vert \int_{\R^n} \Phi \, |D|^{1/2}
\, |u|^2  \, dx\right \vert  \notag  \\ & \le  
c \, ||\Phi \, u||_{L_2} \,  \left \Vert |D|^{1/2} \, u
 \right \Vert_{L_2} +
 c \int_{\R^n} |\Phi| \, | \D_{1/4} u|^2 \, dx 
\notag  \\
& \le  c \, 
||\Phi||_{W_2^{1/2} \to L_2} \,  ||u||^2_{W_2^{1/2}} 
+ c \, \int_{\R^n} |
\Phi| \, |\D_{1/4} J_{1/2} f |^2 \, dx, \notag
\end{align}
where $f = (-1 + \Delta)^{1/4} \, u$. The last integral is bounded 
by:
\begin{align}
 &  \int_{\R^n} | \Phi | \, | J_{1/4} \,\D_{1/4}  \, 
J_{1/4} \,  f|^2 \, dx  \notag \\ & \le  
c \, \int_{\R^n} | \Phi | \,  M  ( \D_{1/4}  \, 
J_{1/4} \,  f) \, | J_{1/2} \,\D_{1/4}  \, 
J_{1/4} \,  f| \, dx
\notag  \\
& \le  c \,  ||M  ( \D_{1/4}  \, J_{1/4} \,  f)||_{L_2} \, 
|| \Phi \, J_{1/2} \,\D_{1/4}  \, 
J_{1/4} \,  f ||_{L_2}
 \notag \\
& \le  c \, || \D_{1/4}  \, J_{1/4} \,  f||_{L_2} \, 
||\Phi||_{W_2^{1/2} \to L_2} \, 
||J_{1/2} \, \D_{1/4}  \, J_{1/4} \, f||_{W_2^{1/2}} \notag \\
& \le  c \, ||\Phi||_{W_2^{1/2} \to L_2} \, ||f||^2_{L_2} = 
c \, ||\Phi||_{W_2^{1/2} \to L_2} \, ||u||^2_{W_2^{1/2}}.
\notag
\end{align}
In the preceding chain of inequalities  we first
applied Hedberg's inequality  (see, e.g.,  \cite{MSh}, Sec. 1.1.3 and 
Sec. 3.1.2):
$$J_{1/4} \, g  \le c \, (M g)^{1/2} \, (J_{1/2} \, g)^{1/2},$$
with $g = |\D_{1/4}  \, J_{1/4} \,  f|,$
 and
then the Hardy-Littlewood maximal inequality for the operator $M$. 
This completes the proof of (\ref{E:2.15}).

To prove the  ``only if'' part of the Theorem, we 
will show that
$$
|| \Phi ||_{W_2^{1/2}(\R^n) \to L_2(\R^n)} \le c \,  
|| \gamma ||_{W_2^{1/2}(\R^n) \to W_2^{-1/2}(\R^n)}.
$$
The proof of this estimate is based on the extension 
 of the distribution $\gamma \in \text{\rm{Mult}} \, 
(W_2^{1/2}(\R^n) \to W_2^{-1/2}(\R^n))$ 
to the higher dimensional Euclidean space, and subsequent application 
of the characterization of 
the class of multipliers $\text{\rm{Mult}} \, 
(W_2^{1}(\R^{n+1}) \to W_2^{-1}(\R^{n+1}))$ obtained by the authors 
in \cite{MV2}.

We denote by $\gamma \otimes \delta$ the  distribution 
on $\R^{n+1}$ 
defined by 
$$\langle  \gamma \otimes \delta, \, u(x,  x_{n+1})   \rangle   
= \langle  \gamma, \, u(x, 0)   \rangle ,$$ 
where $x = (x_1, \ldots, x_n) \in \R^n$,  and
$\delta =  \delta (x_{n+1})$ 
is the delta-function supported on  $x_{n+1} =0$. It is not difficult to 
see that 
$$|| \gamma \otimes \delta ||_{W_2^{1} (\R^{n+1}) 
\to W_2^{-1} (\R^{n+1})} \thicksim  
 || \gamma ||_{W_2^{1/2}(\R^{n}) \to W_2^{-1/2} (\R^{n})}.$$
This follows from the well-known fact that the space of traces on $\R^n$ of
functions in  $W_2^{1}
(\R^{n+1})$  coincides with $W_2^{1/2}(\R^{n})$, with the equivalence
 of 
norms (see, e.g., \cite{MSh}, Sec. 5.1). Indeed, 
for any $U, V \in C^\infty_0 (\R^{n+1})$ let $u(x) = U(x, 0)$ and 
$v(x) = V(x,0)$. Then by the trace estimate mentioned above  
$||u||_{W_2^{1/2}(\R^{n})} \le c \, ||U||_{W_2^{1} (\R^{n+1})},$
and hence
\begin{align}
|\langle  \gamma \otimes \delta, \,
\overline{U}  \,   V \rangle| & = 
|\langle  \gamma, \,  \overline{u} \,  v \rangle|
 \le ||\gamma||_{W_2^{1/2}(\R^{n}) \to W_2^{-1/2} (\R^{n})} 
||u||_{W_2^{1/2}(\R^{n})} \, ||v||_{W_2^{1/2}(\R^{n})} \notag\\
 & \le 
c^2 \, ||\gamma||_{W_2^{1/2}(\R^{n})\to W_2^{-1/2} (\R^{n})}  
\, ||U||_{W_2^{1} (\R^{n+1})} \,
||V||_{W_2^{1} (\R^{n+1})}.\notag
\end{align}
This gives the estimate:
$$|| \gamma \otimes \delta ||_{W_2^{1} (\R^{n+1}) 
\to W_2^{-1} (\R^{n+1})} \le c^2 \, || \gamma ||_{W_2^{1/2}(\R^{n}) \to
W_2^{-1/2} (\R^{n})}.$$ 
The converse inequality (which is not used below) follows similarly 
by extending $u, v \in C^\infty_0(\R^n)$ to 
$U, V \in W_2^1(\R^{n+1})$  with the corresponding estimates of norms.

For the rest of the proof, it will be convenient to introduce  
the notation 
$J_s^{(n+1)} = (-\Delta_{n+1} + 1)^{-s/2}$, $s>0$, 
for the Bessel potential of order $s$ on $\R^{n+1}$; here 
$\Delta_{n+1}$ 
denotes the Laplacian on $\R^{n+1}$.

Now by Theorem 4.2, \cite{MV2} we obtain that 
  $\gamma \otimes \delta \in 
\text{Mult} \, (W_2^{1} (\R^{n+1}) \to W_2^{-1}
(\R^{n+1}))$ if and only if $J_1^{(n+1)} (\gamma \otimes 
\delta) \in  \text{Mult} \, (W_2^{1} (\R^{n+1}) \to L_2(\R^{n+1}))$, 
and
\begin{align}
  || J_1^{(n+1)} (\gamma \otimes 
\delta)||_{W_2^{1} (\R^{n+1}) \to L_2(\R^{n+1})} & \le c \, 
|| \gamma \otimes \delta ||_{W_2^{1} (\R^{n+1}) 
\to W_2^{-1} (\R^{n+1})} \notag \\
& \le  c_1 \, || \gamma ||_{W_2^{1/2}(\R^{n}) \to W_2^{-1/2} (\R^{n})}.
\notag
\end{align}

Next, pick $0<\epsilon <1/2$ and observe that $ J_1^{(n+1)} = (-1 +
\Delta_{n+1})^{1/4 +  \epsilon/2} \, J^{(n+1)}_{\epsilon + 3/2}$.
Using Lemma 2.2 with $l = 1/2 + \epsilon$, $m=1$, and 
$J_{\epsilon + 3/2}^{(n+1)} (\gamma \otimes 
\delta)$ in place of $\gamma$, 
 we deduce:
$$
||J_1^{(n+1)} (\gamma \otimes 
\delta)||_{W_2^{1} (\R^{n+1}) \to L_2 (\R^{n+1})} \thicksim 
  ||J_{\epsilon + 3/2}^{(n+1)} (\gamma \otimes 
\delta)||_{W_2^{1} (\R^{n+1}) \to W_2^{1/2 + \epsilon}(\R^{n+1})}. 
$$
As was proved above, the left-hand side 
of the preceding relation is 
bounded  by 
a constant multiple of 
$|| \gamma ||_{W_2^{1/2}(\R^{n}) \to W_2^{-1/2} (\R^{n})}.$

Thus,
$$||J_{\epsilon + 3/2}^{(n+1)} (\gamma \otimes 
\delta)||_{W_2^{1} (\R^{n+1}) \to W_2^{1/2 + \epsilon}(\R^{n+1})} 
\le 
c \, || \gamma ||_{W_2^{1/2}(\R^{n}) \to W_2^{-1/2}(\R^{n})}.
$$
Passing to the trace on $\R^n = \{ x_{n+1} =0\}$ 
in the multiplier norm on the left-hand side (see \cite{MSh}, Sec. 5.2),
we obtain:
$$
|| \text{Trace} \, J_{\epsilon + 3/2}^{(n+1)} (\gamma \otimes 
\delta) ||_{W_2^{1/2}(\R^{n}) \to W_2^{\epsilon} (\R^{n})}
\le c \, || \gamma ||_{W_2^{1/2}(\R^{n}) \to W_2^{-1/2}(\R^{n})}.
$$
We now observe that 
$$\text{Trace} \, J_{\epsilon + 3/2}^{(n+1)} (\gamma \otimes 
\delta) = \text{const} \, J_{\epsilon + 1/2}^{(n)} (\gamma),$$
which follows immediately by inspecting the corresponding 
Fourier transforms.

In other words,
\begin{equation}\label{E:2.17}
||J_{\epsilon + 1/2}^{(n)} \, \gamma ||_{W_2^{1/2} (\R^{n}) 
\to W_2^{\epsilon}(\R^{n})}  \le 
c \, || \gamma ||_{W_2^{1/2}(\R^{n}) \to W_2^{-1/2}(\R^{n})}.
\end{equation}
From this estimate and Lemma 2.2 with $l = \epsilon$, $m=1/2$, 
and with $\gamma$ replaced by  $J_{\epsilon + 1/2}^{(n)} \gamma$,
  it follows:
\begin{align}
& ||J_{1/2}^{(n)} \, \gamma ||_{W_2^{1/2} (\R^{n}) 
\to L_2 (\R^{n})} = 
||(-\Delta + 1)^{\epsilon/2} J_{\epsilon + 1/2}^{(n)} 
\gamma ||_{W_2^{1/2}(\R^{n})  \to L_2 (\R^{n})}  \notag \\
&  \le  c \, ||J_{\epsilon + 1/2}^{(n)} \gamma ||_{W_2^{1/2}(\R^{n})  \to
W_2^{\epsilon}(\R^{n})} \le C \,  ||\gamma ||_{W_2^{1/2}(\R^{n})  \to
W_2^{-1/2}(\R^{n})}.  \notag
\end{align}
Thus,
$\Phi = J_{1/2}^{(n)} \, \gamma  \in \text{Mult} \, (W_2^{1/2}(\R^{n}) \to
L_2(\R^{n})),$ 
and 
$$
|| \Phi ||_{W_2^{1/2}(\R^n) \to L_2(\R^n)} \le C \,  
|| \gamma ||_{W_2^{1/2}(\R^n) \to W_2^{-1/2}(\R^n)}.
$$
The proof of  Theorem 2.6 is complete.\end{proof}
\vskip13pt 

\rm {

\section{Some corollaries of the form boundedness criterion}
\vskip13pt

Theorem  2.6 proved in Sec. 2, combined with the known  criteria
for nonnegative potentials, yields Theorem II stated in the Introduction.
 In particular, it follows 
that, if $Q \in \D'(\R^n)$,  
and $\Phi = (- \Delta +1)^{-1/4} Q$, then the multiplier defined by  
 $Q$, and hence 
$\mathcal{H} =  \sqrt{- \Delta} + Q$, is a bounded operator from  
 $W^{1/2}_2 (\R^n)$  to $W^{-1/2}_2 (\R^n)$  if 
and only if 
\begin{equation}\label{E:3.1}
\int_e |\Phi (x)|^2 \, dx \le
c   \, \, \text{cap} \, (e, W_2^{1/2} (\R^n)), 
\end{equation}
for every compact set $e\subset \R^n$ such that $\text{diam} \, 
(e) \le 1$. 

Some simpler 
conditions which do not involve capacities are discussed in this 
section.  

The following {\it necessary\/} 
 condition is immediate from  (\ref{E:3.1}) and
the  known estimates of the capacity of the ball in $\R^n$ 
(\cite{MSh}, Sec. 2.1.2). 

\vskip13pt
\noindent {\bf Corollary 3.1.} {\it Suppose $Q \in \D'(\R^n)$, $n\ge 1$. 
Suppose $\mathcal{H} =  \sqrt{- \Delta} + Q: 
\,  W^{1/2}_2 (\R^n) \to W^{-1/2}_2
(\R^n)$ is a bounded operator. Then, for every ball $B_r(a)$ in $\R^n$,
\begin{equation}\label{E:3.2}
\int_{B_r(a)} |\Phi (x)|^2 \, dx \le
c   \,  r^{n-1}, \quad 0<r\le 1, \qquad n \ge 2,
\end{equation}
and 
\begin{equation}\label{E:3.3}
\int_{B_r(a)} |\Phi (x)|^2 \, dx \le
 \frac {c} { \log \frac {2} {r}}, \quad 0<r\le 1, \qquad n =1,
\end{equation}
where the constant does not depend on $a \in \R^n$ and $r$.}
 \vskip13pt

We notice that the class of distributions $Q$ such that 
$\Phi = (- \Delta +1)^{-1/4} Q$ satisfies  (\ref{E:3.2}) 
can be regarded as a Morrey space of  order $-1/2$. 

Combining Theorem II with the Fefferman-Phong condition (\cite{Fef}) 
applied to  $|\Phi|^2$, we arrive at {\it sufficient\/} conditions 
in terms of Morrey spaces of negative order. (Strictly speaking, the 
Fefferman-Phong condition \cite{Fef} was originally 
established for estimates
in the  homogeneous Sobolev space $\dot{W}_2^1$ of order $m=1$. 
However, it can
be  carried over to Sobolev spaces $W_2^m$ for all $0<m\le n/2$. 
See, e.g., \cite{KeS} or \cite{MV1}, p. 98.)

\vskip13pt

\noindent {\bf Corollary 3.2.} 
{\it Suppose $Q \in \D'(\R^n)$, $n\ge 2$. 
Suppose $\Phi = (- \Delta +1)^{-1/4} Q$, and $s>1$.    
Then 
 $\mathcal{H}$ 
 is  a bounded operator
from  $W^{1/2}_2 (\R^n)$ to $W^{-1/2}_2 (\R^n)$ if
\begin{equation}\label{E:3.4}
\int_{B_r(a)} |\Phi(x)|^{2s} \, dx \leq \text{\rm{const}} 
\, r^{n-s}, \quad 0<r\le 1,
\end{equation}
where the constant does not depend on 
$a \in \R^n$ and $r$.} 
\vskip13pt

\noindent {\bf Remark.} 
It is  worth mentioning that  
 condition (\ref{E:3.4}) 
defines a  class of potentials which 
is strictly broader than the (relativistic) Fefferman-Phong class of $Q$ 
such that 
\begin{equation}\label{E:3.5}
\int_{B_r(a)} |Q(x)|^{s} \, dx \leq \text{\rm{const}} 
\, r^{n-s}, \quad 0<r \le 1, \qquad n \ge 2,
\end{equation}
for some $s>1$. 
\vskip13pt

This  follows from the  observation that 
if one replaces $Q$   by $|Q|$ in (\ref{E:3.4}), then obviously 
the resulting 
class defined by:
\begin{equation}\label{E:3.6}
\int_{B_r(a)} (J_{1/2} |Q|)^{2s} \, dx \leq \text{\rm{const}} 
\, r^{n-s}, \quad 0<r \le 1, \qquad n \ge 2,
\end{equation}
becomes  smaller, but still contains some singular measures,  
together with all functions in  the Fefferman-Phong class 
(\ref{E:3.5}). (The latter   was noticed earlier in \cite{MV1}, 
Proposition
3.5.)

A smaller but more conventional class of admissible potentials appears
 when one   
 replaces $\text{cap} \, (e, W_2^{1/2}
(\R^n))$ on the right-hand side of (\ref{E:3.1}) by its lower 
estimate in terms of Lebesgue measure of $e \subset \R^n$. This yields 
the following result (stated as Corollary 1 in the Introduction). 
\vskip13pt

\noindent {\bf Corollary 3.3.} 
{\it Suppose $Q \in \D'(\R^n)$, $n\ge 1$. 
Suppose $\Phi = (- \Delta +1)^{-1/4} Q$.  
Then 
 $\mathcal{H} = 
\sqrt{- \Delta} + Q$ 
 is  a bounded operator
from  $W^{1/2}_2 (\R^n)$ to $W^{-1/2}_2 (\R^n)$ if, 
for every measurable  set $e\subset \R^n$, 
\begin{equation}\label{E:3.7}
\int_e |\Phi (x)|^2 \, dx \le
c   \, |e|^{(n-1)/n},  \quad  \text{\rm{diam}} \, (e) \le 1, \qquad n
\ge 2, \end{equation}
or
\begin{equation}\label{E:3.8}
\int_e |\Phi (x)|^2 \, dx \le
\frac {c} {\log \frac {2}{|e|}}, \quad  \text{\rm{diam}} \,
(e) \le 1,\qquad n =1,
\end{equation}
where the constant $c$ does not depend on $e$.
}
\vskip13pt

We remark that (\ref{E:3.7}), without the extra assumption 
$\text{\rm{diam}} \, (e) \le 1$, is equivalent to 
$\Phi \in L_{2n, \, \infty} (\R^n)$, where $L_{p, \, \infty} (\R^n)$
is the Lorentz (weak $L_p$) space of functions $f$ such that 
$$|\{x \in \R^n: \,  |f(x)| > t\}| \le \frac C {t^p}, \qquad t>0.$$
In particular, (\ref{E:3.7}) holds if 
$\Phi \in L_{2n} (\R^n)$, or equivalently, $Q \in W_{2n}^{-1/2}(\R^n)$.

Furthermore, if $\Phi \in L_{\infty} (\R^n)$, 
then
obviously  (\ref{E:3.7}) holds as well, since 
$$\text{cap} \, (e, W_2^{1/2} (\R^n)) \ge C \, |e|,$$
if $\text{\rm{diam}} \, (e) \le 1$. This leads to the sufficient 
condition $\Phi \in L_{2n} (\R^n) + L_{\infty} (\R^n)$, 
$n \ge 2$.

 It is worth noting that 
(\ref{E:3.7}) defines a  substantially
 broader class of admissible potentials 
 than  the standard (in the relativistic case)  class 
 $Q \in L_{n}(\R^n) +
L_\infty(\R^n)$, $n \ge 2$ (\cite{LL}, Sec. 11.3).  
This is a consequence of the imbedding:
$$L_n (\R^n)\subset W_{2n}^{-1/2} (\R^n), \qquad n \ge 2,$$ 
which  follows   from the classical Sobolev imbedding 
$W_p^{1/2} (\R^n)  \subset L_r(\R^n)$, for  $p=2n/(2n-1)$ and $r=n/(n-1)$, $n
\ge 2$. Indeed, by duality, the latter is equivalent to:
$$L_{n} (\R^n)=L_{r} (\R^n)^* \subset W_{p}^{1/2} (\R^n)^* 
= W_{2n}^{-1/2} (\R^n).$$

Similarly, 
in the one-dimensional case, the class of potentials defined by 
(\ref{E:3.8}) is  wider than the standard class
$L_{1+\epsilon}(\R^1) + L_\infty(\R^1)$, $\epsilon >0$.

It is easy to see that actually  $Q \in  L_n(\R^n)+L_\infty (\R^n)$ 
if $n \ge 2$, or 
$Q \in L_{1+\epsilon}(\R^1)+ L_\infty(\R^1)$ if $n=1$, 
is  sufficient  for the inequality
$$
\int_{\R^n} |u(x)|^2 \, |Q(x)| \, dx
 \leq \text{const} \, ||u||^2_{W_2^{1/2}}, \quad u 
\in C^\infty_0(\R^n),
$$
which 
is  a ``na\"\i ve'' version
of  (\ref{E:1.2}) where  $Q$ is replaced by $|Q|$.

\bigskip

\end{document}